\documentclass{aastex631}

\shorttitle{Changes in GRS 1915+105}
\shortauthors{Rodr\'\i guez \& Mirabel}

\graphicspath{{./}{figures/}}

\shorttitle{Changes in GRS 1915+105}
\shortauthors{Rodr\'iguez \& Mirabel}

\begin{document}

\title{An Unusual Change in the Radio Jets of GRS 1915+105}

\author[0000-0003-2737-5681]{Luis F. Rodr\'{\i}guez}
\affiliation{Instituto de Radioastronom\'{\i}a y Astrof\'{\i}sica\\
Universidad Nacional Aut\'onoma de M\'exico, Apdo. Postal 3-72, Morelia, Michoac\'an 58089, Mexico}
\affiliation{Mesoamerican Center for Theoretical Physics\\
Universidad Aut\'onoma de Chiapas, Tuxtla Guti\'errez, Chiapas 29050, Mexico}
\author[0000-0002-3210-6307]{I. F\'elix Mirabel}
\affiliation{D\'epartement d’Astrophysique-IRFU-CEA \\ Universit\'e Paris-Saclay, France}
\affiliation{Instituto de Astronom\'{\i}a y F\'{\i}sica del Espacio (IAFE)\\
CONICET-Universidad de Buenos Aires, C1428 Buenos Aires, Argentina}

\begin{abstract}

We compare Very Large Array observations of GRS 1915+105 made in 1994 and 2023, with nearly three decades of difference.
The source has experienced intriguing major changes. The position angle of the bipolar ejecta in the
plane of the sky has increased counterclockwise by 24$^\circ$. The inclination angle of the 
flow with respect to the line of sight has increased by 17$^\circ$. Both sets of observations
show the temporal quasi-sinusoidal 
radio oscillations previously reported for other epochs. However, the 2023 oscillations
are faster than ever before, with a period of about 8 minutes as opposed to the periods in the range of 20 to 40 minutes
observed in previous epochs.
Analysis of GRS 1915+105 images over the years
suggest that the observed changes took place within a year or less. Our analysis indicates
that during 2023 the plane of the accretion disk was aligned with the line of sight, which may explain the deep X-ray obscured state and the high mid-infrared luminosity observed with JWST in that epoch. 
More recent 2024 observations
show that the position angle of the ejecta has returned to its historic values. 
Future 
monitoring of the time evolution of the source may clarify the cause of these remarkable changes.

\end{abstract}



\keywords{Radio continuum emission (1340) -- Relativistic jets(1390) -- X-ray binary stars(1811)}

\section{Introduction} \label{sec:intro}

The hard X-ray transient GRS 1915+105 was discovered in 1992 by the WATCH all sky X-ray monitor on board of the GRANAT space observatory (Castro-Tirado et al. 1994). In 1993 Finoguenov et al. (1994) localized the high energy source inside a $3\rlap'.5$ error radius of the SIGMA Gamma-Ray Telescope on GRANAT. Following observations with the Very Large Array (VLA) of NRAO and the infrared camera IRAC2 on the 2.2-m telescope of ESO, a time variable radio/infrared counterpart of the high energy source was identified in 1993 by Mirabel et al. (1994) close to the galactic plane, beyond $\sim$20 magnitudes of optical absorption. 
This radio/infrared counterpart was monitored at several radio frequencies, with the VLA and Nançay radio telescopes in December 1993 to April 1994 by Rodríguez et al. (1995). The VLA observations of the brightest outburst revealed that GRS 1915+105 produces double-sided relativistic ejections of plasma clouds that appear to have superluminal transverse motions (Mirabel \& Rodríguez 1994). Super-luminal motions had been observed before only for radio-emitting components in a number of distant quasars and active galactic nuclei. Therefore, GRS 1915+105 became the first example of the superluminal phenomenon in the Milky Way (Mirabel \& Rodríguez 1999). Additional relativistic ejection events in GRS 1915+105, with about similar radio parameters as in the 1994 events, repeated in 1995 and 1997, during this historical period of bright average X-ray flux (Rodríguez \& Mirabel 1999). Over the years ejection 
events have been observed without an evident periodicity.

During the 1995-1997 period Pooley and Fender (1997), surveyed the source at 15 GHz with the Ryle telescope finding that the radio emission varied in strong correlation with the X-ray emission as recorded by the RXTE all-sky monitor, exhibiting quasi-periodic oscillations with periods in the range of 20-40 min. However, after more than 20 years of intense X-ray activity, in the year 2018 GRS 1915+105 initiated a decline in X-ray flux reaching a few years later a state of almost complete “obscured X-rays”, but with activity at radio waves. To solve this puzzle, it has been proposed 
that intrinsic obscuration of the X-rays is present
(Miller et al. 2020; Motta et al. 2021; Balakrishnan et al. 2021; S{\'a}nchez-Sierras et al. 2023). Mid-infrared spectral timing 
observations with the JWST during this X-ray obscured state were made in 2023 (Gandhi et al. 2015), with the source exhibiting 
exceeding past infrared levels by about a factor of 10. The unusual changes in the radio jets of GRS 1915+105 reported here were 
found in archive 
VLA observations during the X-ray deep obscured state of the source, carried out 116 days after the JWST 
observations of an unusually high Mid-Infrared flux from the source. 

New VLA observations of GRS 1915+105 in the same array configuration were made in 30 September + October 01, 2023
(project 23B-314, PI: S.E. Motta), soon after Mid-Infrared observations on 6 June 2023 with MIRI on the JWST and AMI-LA at 15 GHz, when the source was in an infrared/radio-bright state, but X-ray-obscured state 
(Miller et al. 2020; Balakrishnan et al. 2021; Gandhi et. al. 2025, and references therein). 
Here we report the observation of a significant change in the radio parameters of the jets from the black hole (BH)-XRB GRS 1915+105 between these two epochs almost 30 years apart. In section 2 the observations of two epochs are described, in section 3 they are compared, and in section 4 we discuss the time evolution of the changes.  Finally, the discussion and conclusions are presented in
sections 5 and 6.

\section{Observations} 
\label{sec:observations}

We have used the observations described in Table 1 from the archives of the 
VLA of NRAO\footnote{The National 
Radio Astronomy Observatory is a facility of the National Science Foundation operated
under cooperative agreement by Associated Universities, Inc.}. In both epochs the array was in its highest
angular resolution A configuration
and the amplitude calibrator was J1331+305 (3C286).

The 1994 data were analyzed in the standard manner using the
AIPS (Astronomical Image Processing System) package. The 2023 data 
were calibrated using the 
CASA (Common Astronomy Software Applications;  McMullin et al. 2007) package of NRAO and
the pipeline provided for VLA\footnote{https://science.nrao.edu/facilities/vla/data-processing/pipeline} observations. 
The 2023 data were obtained at the C-band (4-8 GHz) and X-band (8-12 GHz) receivers and to obtain a better signal-to-noise ratio the data were concatenated to obtain a total coverage of 4 to 12 GHz, which gives a central frequency of 8.0 GHz and a bandwidth of 8 GHz. We made images for both epochs using a robust weighting (Briggs 1995) of 0 to
optimize the compromise between angular resolution and sensitivity.  All images were also corrected for the primary beam
response.

 
 In Figure 1 we show contour images of GRS 1915+105 for both epochs (left and center panels).
 In both cases the source shows three components: a central core believed to trace the position of the X-ray binary and a northern and southern lobe tracing recent ejecta from the X-ray binary. In the case of the 1994 image we know that this last
 assumption is correct because Mirabel \& Rodríguez (1994) followed the motion of the lobes in seven epochs over slightly more than a
 month. Unfortunately, in the case of 2023 only one epoch is available. The 2023 image has a much larger noise than expected,
 a result of the large and fast variability that the core presented at that epoch (Rodríguez  \& Mirabel, in preparation). 
 
In addition to the two images just discussed we present in the right panel of Figure 1 a more recent VLA image, taken on 2024 May 25. 
These data were taken at a center frequency of 5.5 GHz with a bandwidth of 2.0 GHz and using
3C286 as amplitude calibrator and J1922+1530 as gain caiibrator. These data were calibrated in
the same manner as the data from 2023. This last image has 
 poorer angular resolution since it was taken in the B configuration as part of project 24A-474 (PI: S.E. Motta).
 Fortunately, the outflow was relatively extended in that epoch and it was possible to determine that the position angle 
 of the ejecta was 
 $152^\circ \pm 1^\circ$, indicating a return to its historic values. In this image, given
 the lower angular resolution, we cannot isolate the emission of the core.
    

\begin{deluxetable*}{ccccccccc}
\tablenum{1}
\tablecaption{Parameters of the VLA Observations}
\tablewidth{900pt}
\tabletypesize{\scriptsize}
\tablehead{
 \colhead{}    & \colhead{} & \colhead{Gain} & \colhead{Frequency} & \colhead{Bandwidth} & 
 \multicolumn{2}{c}{Position of core (J2000)} & \colhead{PA}
   & \colhead{i}  \\
\colhead{Epoch} &  \colhead{Project} & \colhead{Calibrator} & \colhead{(GHz)} 
 & \colhead{(GHz)} &   RA($19^h~15^m$) & DEC($+10^\circ~56'$)
      & \colhead{($^\circ$)} & \colhead{($^\circ$)} }   \decimalcolnumbers
       \startdata
      1994 Apr 23 & AR277 & J1925+211 & 8.4 &  
      0.1  & $11\rlap.{^s}551$ & $44\rlap.{''}78$ &  149.7$\pm$0.7 &  70$\pm$2 \\ 
     2023 Sep 30+Oct 01 & 23B-314 & J1924+1540 & 8.0 & 8.0 &
      $11\rlap.{^s}544$ &  $44\rlap.{''}61$  & 174.3$\pm$0.5& 87$\pm$3  \\
     \enddata  
\tablecomments{The positions are assumed to have a precision of $0\rlap.{''}01$ (Smol{\v{c}}i{\'c} et al. 2017; Rodriguez et al.
2024). PA = position angle of the ejecta
in the plane of the sly. i = inclination angle of the southern flow with respect to the line of sight.}
\label{tab:param}
\end{deluxetable*}

\begin{figure*}[!t]
\vskip-1.0cm
\hskip-0.1cm\includegraphics[width=1.0\linewidth]{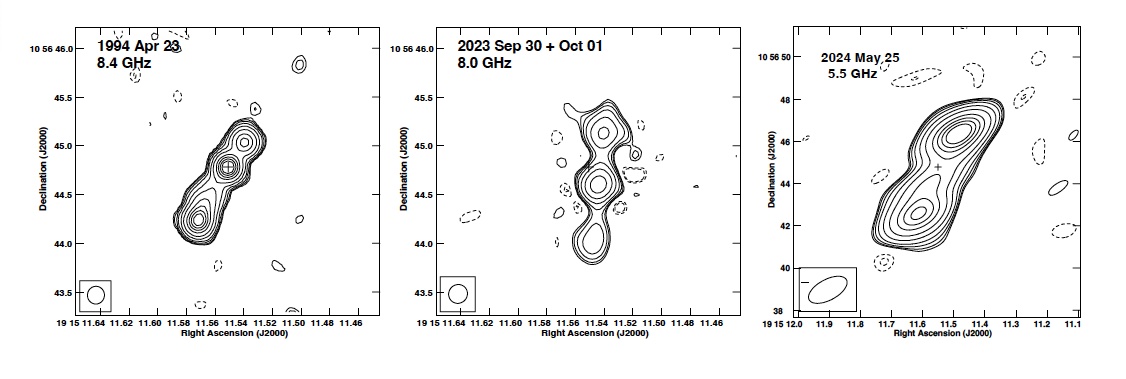}
\vskip-0.7cm
\caption{Very Large Array contour images of GRS 1915+105 for epochs 1994.310 (left), 2023.748 (center) and 
2024.399 (right).
Contours 
are -4, -3, 3, 4,  6, 10, 20, 40, 80, 120, 160 and 200
times the rms noise of the image, 120  $\mu$Jy beam$^{-1}$ for the 1994.310 epoch, 
660 $\mu$Jy beam$^{-1}$ for the 2023.748 epoch and 
10 $\mu$Jy beam$^{-1}$ for the 2024.399 epoch. The non-Gaussian noise pattern around the core
for the 2023 image is due to its fast variability during these observations (see Figure 2). 
The synthesized beams,  shown in the bottom left corner of each image, are 
($0 \rlap.{"}18 \times 0\rlap.{"}19; -5\rlap.^\circ2$), ($0 \rlap.{"}20 \times 0\rlap.{"}19; -43\rlap.^\circ2$)
and ($2 \rlap.{"}01 \times 0\rlap.{"}98; -61\rlap.^\circ9$)
for the 1994.310, 2023.748 and 2024.399 epochs, respectively. The cross marks the 1994.310 position of the central component,
assumed to trace the position of the X-ray binary.
 }
\label{fig:3epochs}
\end{figure*}

\section{Comparison of the images}
\label{sec:comparison}

\subsection{Proper motions of the source}

In Figure 1 it is evident that the core of the radio source, tracing the X-ray binary, 
has displaced between 1994 and 2023. 
Over the 29.44 yr between these observations this corresponds to a proper motion of $\mu_{RA}$ = --3.46 $\pm$ 0.48 
mas yr$^{-1}$, $\mu_{DEC}$ = --5.94 $\pm$ 0.48 mas yr$^{-1}$.
These values are consistent at the 1-$\sigma$ level with the highly accurate values of 
$\mu_{RA}$ = --3.14 $\pm$ 0.03 mas yr$^{-1}$, $\mu_{DEC}$ = --6.23 $\pm$ 0.04 mas yr$^{-1}$ obtained from the
variance-weighted average of Very Long Baseline Array observations of Dhawan et al. (2007) and Reid et al. (2014).
Then this displacement is as expected from previous observations.

\subsection{Position angle of the ejecta}

This position angle was calculated from the positions of the lobes. The values are listed in Table 1. Over the time period of the 
observations the position angle has increased in the counterclockwise sense by $\Delta PA = 24\rlap.^\circ 6 \pm 0\rlap.^\circ 9$.

\subsection{Inclination angle of the ejecta}

The inclination angle of the southern jet with respect to the line of sight, $i$,  was estimated from
the relative proper motions of the approaching and receding condensations
to be $70^\circ \pm 2^\circ$ 
by Mirabel \& Rodriguez (1994), as included in Table 1. Using the compilation of accurate proper-motion
measurements from Miller-Jones et al. (2007), Reid \& Miller-Jones (2023) derive a weighted mean inclination angle of
$64^\circ \pm 4^\circ$, consistent with the value of Mirabel \& Rodriguez (1994). Since the weighted mean inclination angle
of Reid \& Miller-Jones (2023) is an average over 1994 to 2006, the agreement between both measurements suggests that
there were no major changes in this parameter over that period of time. Reid \& Miller-Jones (2023) also conclude that,
for different epochs, the true jet velocity is in the range $\beta = 0.80 \pm 0.12$. We have assumed that this value of
$\beta$ is constant, an assumption supported by the consistent proper motions measured over many years 
(e.g. Miller-Jones et al. 2007). 

Since for 2023 we only have one epoch, we assumed that both lobes seen in Figure 1 were ejected at the same time.
In this case one of the equations that describe the phenomenon (Mirabel \& Rodriguez 1994) simplifies to:

$$cos(i) = {{(\Delta \theta_a - \Delta \theta_r)} \over {\beta (\Delta \theta_a + \Delta \theta_r)}},$$

\noindent where $\Delta \theta_a$ and $\Delta \theta_r$ are the angular displacements of the approaching and receding lobes,
respectively, from the central source. For the 2023 data we obtain $\Delta \theta_a = 0\rlap.{''}57 \pm 0\rlap.{''}02$ 
and $\Delta \theta_r = 0\rlap.{''}53 \pm 0\rlap.{''}02$. These values imply $i = 87^\circ\pm 3^\circ$, that is, with the outflow axis
practically perpendicular to the line of sight. This result is insensitive to the exact value of $\beta$. Adopting a generous range of values 
from 0.50 to 0.99 we find that the derived inclination angle ranges from $86^\circ$ to $88^\circ$. We conclude that 
between 1994 and 2023 $i$ has increased by $\Delta i = 17^\circ\pm 4^\circ$.

\subsection{Total angular change}

The addition of the two angular displacements (in the plane of the sky and along the line of sight) is given by the first spherical law of cosines:

$$cos(\Delta \phi_T) = cos(\Delta PA) \cdot cos(\Delta i),$$

\noindent where $\Delta \phi_T$ is the total angular change of the outflow axis. Substituting the values for $\Delta PA$
and $\Delta i$ given above, we obtain

$$\Delta \phi_T = 30^\circ\pm 2^\circ.$$

\subsection{Periodic variability in the radio emission from the core}

The central source of GRS 1915+105, associated
with the binary system, is known to exhibit quasisinusoidal oscillations in its centimeter flux density (Fender et al. 1997; Rodríguez \& Mirabel 1997; Pooley \& Fender 1997). Characteristically, these quasi-periodic oscillations have periods in the range 20-40 min.
To obtain the flux density of the core as a function of time we cleaned and subtracted the clean components of the lobes using the CASA tasks TCLEAN and UVSUB. Finally, we centered the core in the (u,v) plane with the task FIXVIS and
obtained the flux densities with the task PLOTMS.

In Figure 2 (left panel) we show the oscillations detected for 1994 April 23 at 8.4 GHz, that have a period of 24-min.
In contrast, the oscillations detected for 2023 Sep 30 at 9 GHz, shown in Figure 2 (right panel), have a period of only 8-min.
It is suggestive to associate the change in the period of the radio flux density oscillations 
with the change in the orientation of the jets in 2023. We already noted that the period of the radio flux density of 24-min 
in 1994 was reduced to a period of 8-min in 2023 (Fig. 2). To analyze the association between the changes in period 
with jet orientation we assume that the jet direction is perpendicular to the binary orbital plane, which implies that the inner 
accretion disk is coplanar with the binary orbit. This is expected given the old age of GRS 1915+105, which has a 
longer orbital period and a more evolved stripped-giant donor star than the prototype black-hole X-ray transient V404 Cygni, 
which has an estimated age of 4$\pm$1 billion years (Burdge et al. 2024). From direct measurement of the orbital period in 2000 April-August (Greiner et al. 2001) and between 2010 June and 2011 August (Steeghs et al. 2013), it is concluded that the binary orbit in those historic epochs was circular. 

We will assume that the observed periodicities arise from the fact that most of the emission is coming from a characteristic radius in the accretion disk. We further assume the same binary orbital parameters in 1994 than those in the years 2000, 2010 and 2011. 
Using Kepler’s third law it is inferred that the semi-major axis of the accretion disk in 2023 was reduced by a factor of 0.48 
relative to those historic epochs. Finally, assuming that the mass is conserved, this reduction
in radius implies an enhancement of the surface gas density of the accretion disk by a factor of $0.48^{-2} \simeq$
4.3, possibly implying an increase of the accretion rate comparable to this factor. 
Although the thickness, shape and orientations of the accretion flow in 2023 are not constrained, a large 
increase in accretion rate could cause important changes in the parameters of the system with respect to 1994, 
potentially increasing the accretion disk eccentricity and originating a transient misalignment between the angular momentum 
of the accretion flow and the black hole spin. In this context, the increase of accretion rate in 2023 could have induced 
an enhancement of the intrinsic X-ray, MIR and radio luminosities of the source.
However, it is presently unknown if a change in accretion rate could 
produce the required dramatic effects.

In fact, the VLA observations shown in the central panel of Figure 1 and right panel of Figure 2 were made in 30 Sept. + Oct. 01, 2023, after a MIR-Radio-bright and X-ray-obscured state observed in June 2023 with MIRI on board the JWST. These mid-infrared observations showed that at that time the source was exceeding past infrared levels by about a factor of 10, 
implying high-density gas column densities of $N_H \geq 3 \times 10^{23}~cm^{-2}$ while the source was undergoing radio burst 
activity (Gandhi et al. 2025, Fig. 1 and references therein). In contrast, the observed X-ray fluxes at that time were much fainter than the historical average of the source's persistent X-ray `obscured' state. However, the transitions of polycyclic aromatic hydrocarbon molecules (PAHs) associated with dust, that were observed several times in 2004 and 2006 with the Spitzer Space Telescope (Rahoui et al. 2010), were absent in the spectra with MIRI on the JWST (Gandhi et al. 2025). This absence of PAHs in 2023 is likely due to heating of the dust and molecular ionization by strong X-ray/UV radiation from the inner parts of the accretion disk.  

Much faster radio oscillations, with periods of about 0.2 seconds, have been reported by Tian et al. (2023). At present
the nature and possible relation of both types of oscillations is not understood.

\begin{figure*}[!t]
\vskip-1.0cm
\hskip-0.5cm\includegraphics[width=0.45\linewidth]{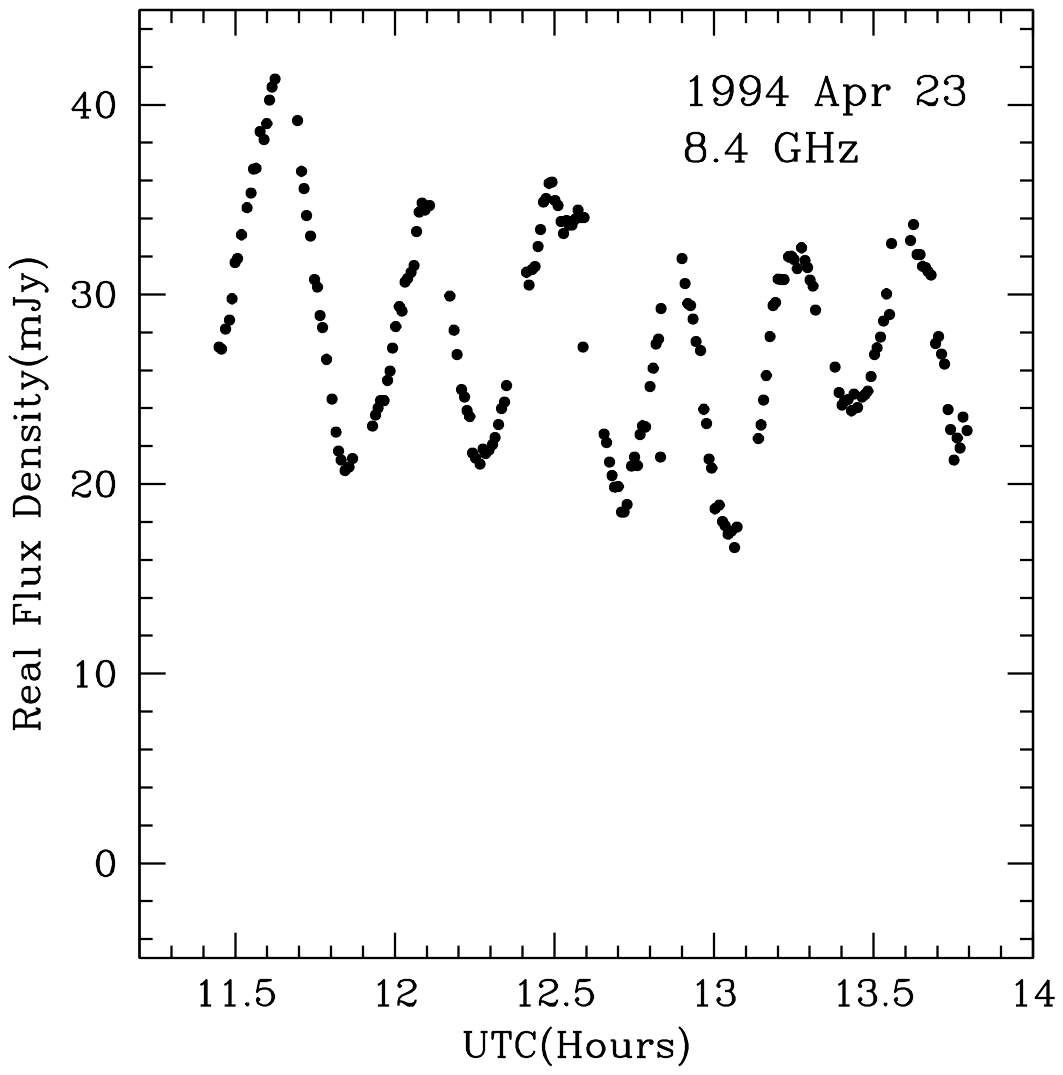}\hskip-1.2cm\includegraphics[width=0.45\linewidth]
{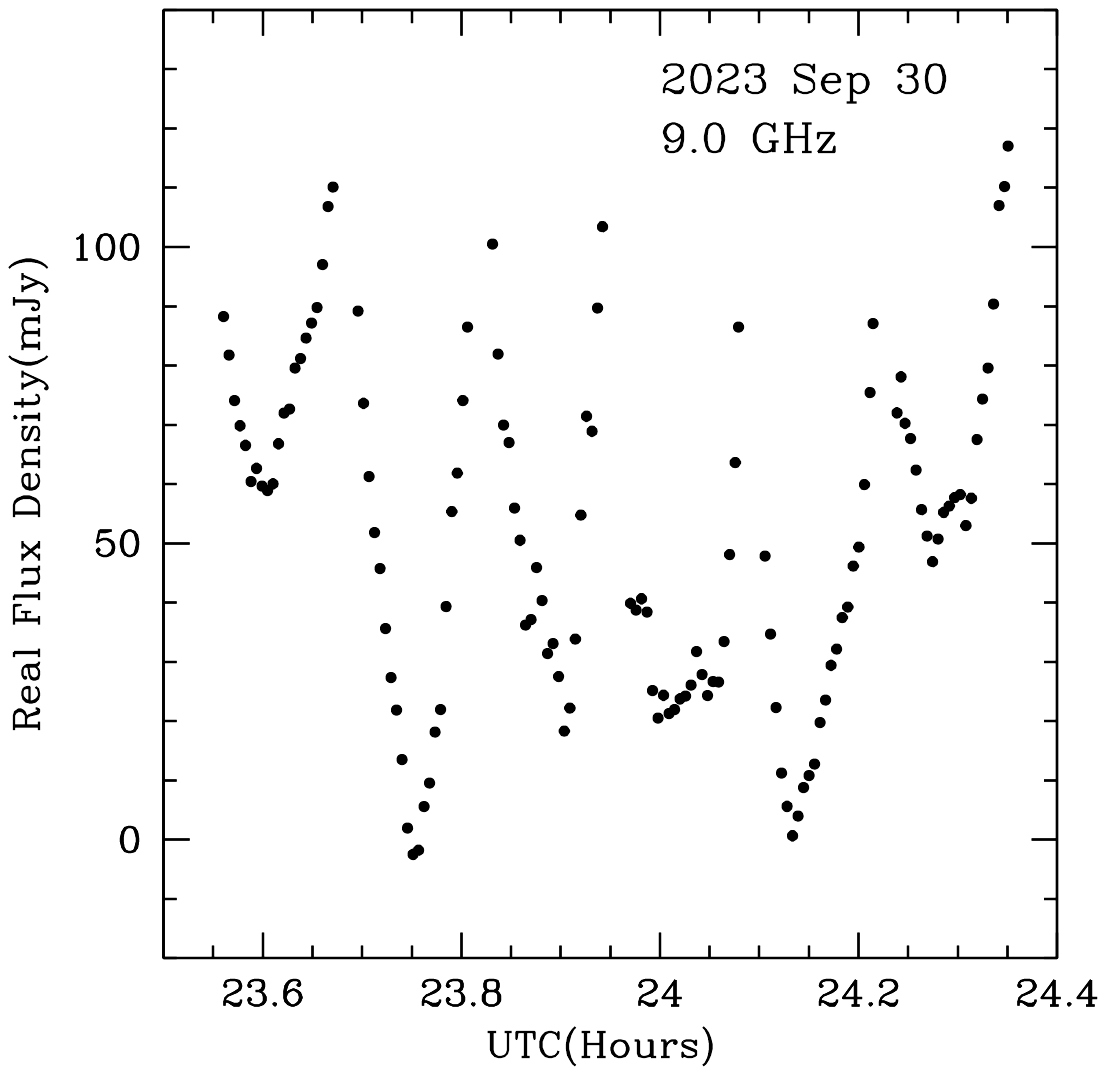}
\vskip-1.9cm
\caption{Flux density of the core as a function of Universal Time Coordinated (UTC) 
for  1994 April 23 (left) and 2023 Sep 30 (right). Note the change in 
period from 24 to 8 minutes}
\label{fig:2epochs}
\end{figure*}

\section{The Time Evolution of the Changes}

Did the position angle changes between 1994 and 2023 took place slowly or suddenly?

To address this key question we compiled data from the literature and reduced and analyzed VLA archive data to
study the time behavior of the position angle with time. The position angle values obtained are listed in Table 2 and plotted in Figure 3. 
When possible, these position angles were accurately obtained from the positions of the antisymmetric lobes.
In the case of the VLBA observations and some of the VLA observations the lobes were not detected
and the position angle was obtained from 
a Gaussian ellipsoid fit to the central component. The position angles determined by this last method have much larger errors.
In Table 3 we give the positions of the lobes for the last three epochs listed in Table 2, from which accurate position angles
(listed in Table 2) can be obtained.

Not taking into account the observation of 2023 Sep 30+Oct 01, we obtain a position angle weighted mean and rms of 
$147^\circ \pm ~8^\circ$ from the remaining 18 epochs. In Figure 3 we plot this mean and the $\pm 1-\sigma$ range.
We then find that the 2023 Sep 30+Oct 01 position angle differs by 3.4$\sigma$ from the mean, supporting a significant change.
Furthermore, from Figure 3 we can see that the two position angles measured in 2024, after the 2023 departure, are again consistent with the mean. We can then conclude that the departure from the characteristic value and the return to it took place in about a year or less. This behavior differs strongly from the smooth and continuous change in the outflow axis produced by precession, as
observed for example in SS 433 (Margon 1984).
The dotted lines in Figure 3 may trace the upper limits of a possible regular precession of the direction of the jets of GRS 1915+105, 
with a semi-angle, within the uncertainties, similar to that of $\sim10^\circ$ observed in SS433 (Margon \& Anderson 1989).
However, the presence of a significant regular precession is at odds with the consistent proper motions measured over many years (e.g. Miller-Jones et al. 2007). To test this issue we did a Lomb-Scargle periodogram (Press et al. 1992) using the data 
of Table 2 (excluding the 2023 data point), failing to find a significant periodicity and favoring
the hypothesis of a lack of a significant regular precession.

\begin{deluxetable*}{ccccc}
\tablenum{2}
\tablecaption{Position angle of the GRS 1915+105 outflow in time}
\tablewidth{900pt}
\tabletypesize{\scriptsize}
\tablehead{
\colhead{Epoch} &  \colhead{MJD} &  \colhead{PA($^\circ$)} & \colhead{Instrument} 
      & \colhead{Reference}  }   
      \decimalcolnumbers
       \startdata
      1994 Jan 29   &   49381 & 159$\pm$8   &      VLA   &         Rodr{\'\i}guez \& Mirabel(1999)  \\ 
      1994 Feb 19   &   49402 & 157$\pm$6      &    " & " \\
1994 Mar 19  &    49430  & 149$\pm$4     &      " & "  \\
1994 Apr 21   &   49463  & 147$\pm$6     &      "         &         "  \\
1994 Apr 23   &    49465  & 149.7$\pm$0.7 & " & This paper \\
1995 Aug 10 &  49939  & 140$\pm$10    &     "         &         "  \\
1997 Oct 23  &  50744   &  157$\pm$2    &     VLBA  &  Dhawan et al.(2000) \\
1997 Oct 31 &   50752  &   133$\pm$3    &        "      &       "  \\
1997 Oct 31   &  50752  &    143$\pm$4    &        "         &     "    \\
1997 Nov 06   &  50758 &  142$\pm$2   &  MERLIN &   Fender et al.(1999)  \\
1998 Apr 11  &  50914  &   155$\pm$2           &            VLBA  &  Dhawan et al.(2000) \\
1998 May 02  &  50935    &   154$\pm$4        &                 "      &      "  \\
1998 May 02   &  50935  &  145$\pm$6        &                     "     &      " \\
2013 May  24 &  56436   &  130$\pm$1  &  VLBA   &     Reid et al.(2014) \\
2015 Mar 05 &  57086  & 140$\pm$8  & VLA  &  15A-460 \\
2023 May 23   &  60088    &    138$\pm$9    &   " & SS192168 \\
2023 Sep 30+Oct 01 & 60217.5   &    174.3$\pm$0.5  & "     &   23B-314  \\
2024 May 25  &  60455    &   152$\pm$1  & "  & 24A-474 \\
2024 Jun 23  & 60484    &     149$\pm$1   &   "  & 24A-474 \\      
     \enddata  
\tablecomments{MJD = Modified Julian Date. PA = position angle of the ejecta
in the plane of the sky.. VLBA = Very Long Baseline Array. MERLIN = Multi-Element Radio Linked Interferometer Network.
The MERLIN observations are the average of 10 epochs observed across 11 days.}
\label{tab:patime}
\end{deluxetable*}

\begin{figure*}
\vskip-2.0cm
\includegraphics[width=0.8\columnwidth]{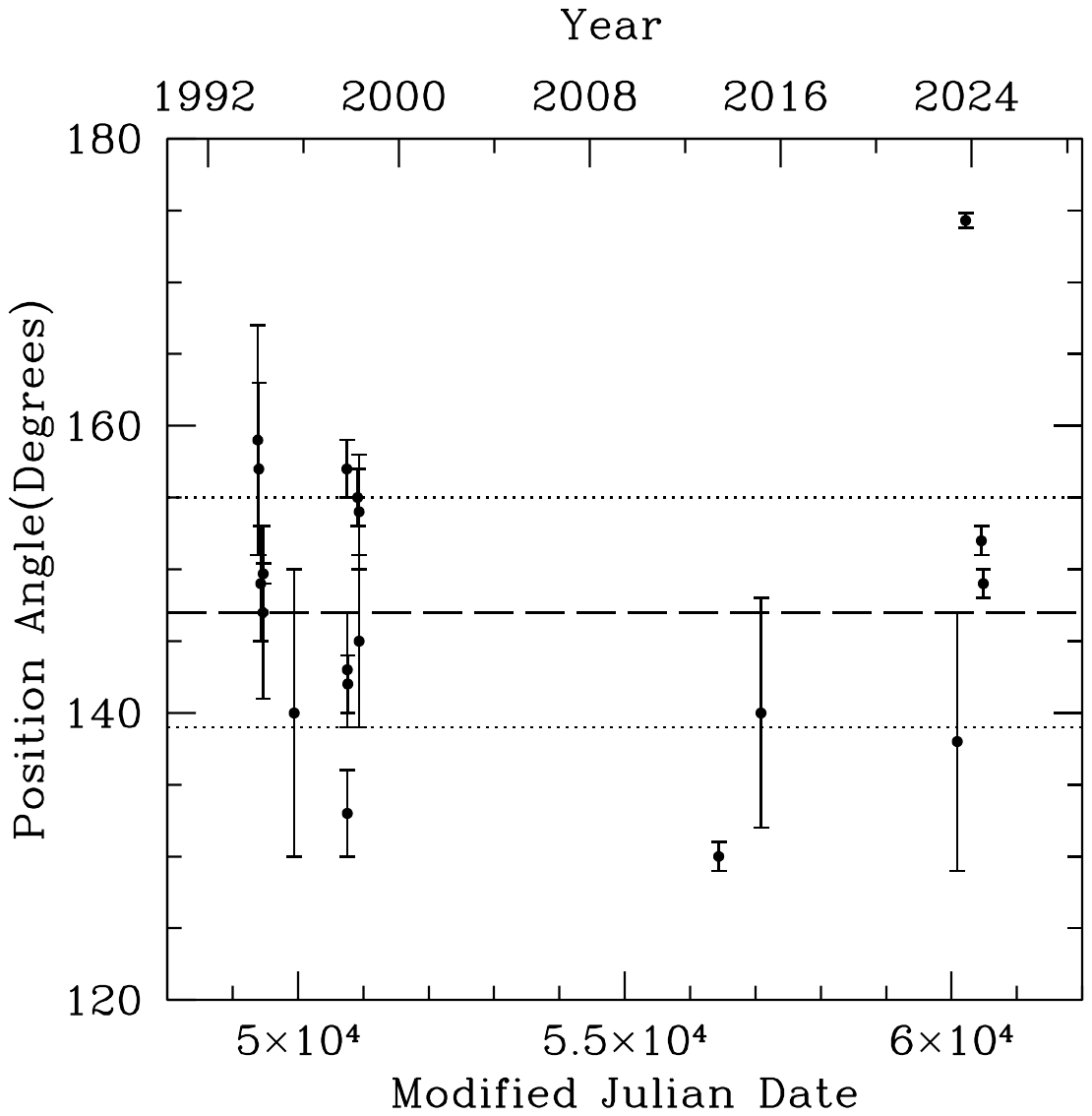}
\vskip-3.4cm
\caption{Position angle of the GRS 1915+105 ejecta as a function of Modified Julian Date (MJD; bottom line)
and year (top line). The dashed line indicates 
the weighted mean for all the points, with the exception of the 2023 Sep 30+Oct 01 data. The weighted $\pm1-\sigma$ range is indicated with the dotted lines.
}
\label{fig:grsPA}
\end{figure*}

\begin{deluxetable*}{ccccccccc}
\tablenum{3}
\tablecaption{Positions of the Lobes for the Last Three Epochs of Table 2}
\tablewidth{900pt}
\tabletypesize{\scriptsize}
\tablehead{
 \colhead{}    & \colhead{} &  \colhead{VLA} & \colhead{Frequency} & \colhead{Bandwidth} & 
 \multicolumn{2}{c}{Position of Northern Lobe (J2000)} & \multicolumn{2}{c}{Position of Southern Lobe (J2000)} \\
\colhead{Epoch} &  \colhead{Project} & \colhead{Configuration} & \colhead{(GHz)} 
 & \colhead{(GHz)} &   RA($19^h~15^m$) & DEC($+10^\circ~56'$)
&  RA($19^h~15^m$) & DEC($+10^\circ~56'$)}
       \startdata 
     2023 Sep 30+Oct 01 & 23B-314 & A & 8.0 & 8.0 &
    $11\rlap.{^s}5402\pm0\rlap.{^s}0003$  & $45\rlap.{''}135\pm0\rlap.{''}006$ & $11\rlap.{^s}5476\pm0\rlap.{^s}0004$  
    & $44\rlap.{''}040\pm0\rlap.{''}010$  \\
           2024 May 25 & 24A-474 & B & 5.5 & 2.0 &
      $11\rlap.{^s}4682\pm0\rlap.{^s}0024$  &  $46\rlap.{''}298 \pm 0\rlap.{''}027$  & $11\rlap.{^s}6104\pm0.{^s}0037$ 
      &  $42\rlap.{''}808\pm0\rlap.{''}060$ \\
           2024 Jun 23 & 24A-474 & B & 5.5 & 2.0 &  $11\rlap.{^s}4829\pm0\rlap.{^s}0021$ &  $46\rlap.{''}076\pm0\rlap.{''}073$ & 
           $11\rlap.{^s}6229 \pm0\rlap.{^s}0027$ & 
      $42\rlap.{''}643\pm0\rlap.{''}076$  \\         
      \enddata  
\label{tab:param}
\end{deluxetable*}

\section{Discussion}


Theoretically, the X-ray weakness of BHs such as those expected to exist 
in low-luminosity active galactic nuclei at z$>$4
could be explained by mildly super-Eddington accretion onto slowly spinning BHs 
(e.g. Pacucci and Narayan 2024). However, this mechanism does not
work for GRS 1015+105 because it has been concluded from the analysis of X-ray spectral observations that the compact 
object in this binary is a nearly extreme rapidly rotating Kerr BH (McClintock et al. 2006), and it is not clear how mildly super-
Eddington accretion could have slowed down drastically such rapidly rotating BH. Furthermore, Figure 1 shows that after the unusual 
change of orientation of the radio jets in 2023, in 2024 they
returned to the range of their historical values, which indicates that the unusual orientation of the jets in 2023 was a transient event, 
rather than a change in the BH-spin.  This behavior is in agreement with the Bardeen-Petterson (1975) effect that implies that
the axis of a tilted disk around a Kerr
black hole will transition to be aligned with the spin of the black hole.               

The source’s persistent X-ray ‘obscured’ and radio bright flare state since 2018 (Motta et al. 2021), suggests that the obscured X-ray state is due to strong intrinsic absorption of the X-rays, probably by Compton scattering in the high-density gas column densities of 
$N_H > 3 \times 10^{23}~ cm^{-2}$ 
observed in the mid-IR with JWST. These large column densities suggest accretion rates close to or above Eddington by a thick disk. 

Rapid changes in the orientation of relativistic jets were observed with the VLBA on time scales of minutes to hours, by Miller-Jones et al. (2019) in the microquasar V404 Cygni, which as GRS 1915+105 is a BH-LMXB, with analogous intrinsic properties, that also undergoes periods of persistent X-ray ‘obscured’ states. Miller-Jones et al. (2019) propose that the changes in the orientation of the jets in V404 Cygni are due to precession of the jets, by frame dragging when the orbital plane of the companion star is misaligned with the BH spin axis (e.g. Stella et. al. 1998). However, the binary orbital inclination has not been independently constrained in GRS 1915+105. 

As shown in subsection 3.3, the inclination angle i of the southern jet of GRS 1915+105 with respect to the line of sight between 1994 and 2023 increased by $17^\circ \pm 4^\circ$, to i = $87^\circ \pm 3^\circ$, which implies that the inclination angle of the jets axis with the plane of the sky in 2023 is of $3^\circ \pm 3^\circ$. Assuming that the jet axis is perpendicular to the accretion disk, the line of sight to the X-ray source in 2023 is along the plane of the accretion disk. In this particular configuration, relativistic boosting and de-boosting should not be significant, and as expected, the radio brightness of the Southern and Northern lobes in the observations of 2023 are about the same, as shown in Figure 1. 

Therefore, the line of sight along the plane of the thick accretion disk could explain the very deep X-ray obscuration of GRS 1915+105 in 2023. This geometry could also explain the unprecedented MIR-bright state, and the mid-IR emission-line lag relative to the underlying continuum, that is consistent with the characteristic timescales of the outer accretion disk, observed in that epoch with MIRI on board JWST (Gandhi et al. 2025). 

We already noted that the abrupt change in the outflow axis of GRS 1915+105 differs from the smooth changes 
produced by a continuous precession. What may have cause that unusual change? In fact, the
formation of BH-LMXBs has been a long-standing question. Theoretical work suggested that hierarchical triples might be key to form BH-LMXBs (Naoz et al. 2016). Support to that theoretical prediction is the recent discovery by Gaia of a hierarchical triple with a tertiary companion, thousands of astronomical units away from the inner binary in V404 Cygni (Burdge et al. 2024). 
In this context, the unusual changes in the GRS 1915+105 system could be due to the presence of a third component still unidentified. A third component, if present in a high eccentricity orbit, could perturb the inner compact binary periodically.
On the other hand, in BH-LMXBs strongly variable high mass transfer rates by Roche lobe overflow may take place due to nuclear evolution of the donor, when it is a low-mass giant star (King et al. 1996). These events could induce unusual impacts on the accretion disk. These hypotheses will be discussed for the case of GRS 1915+105 in work in progress.

\section{Conclusions}

1) We analyzed archive VLA observations of GRS 1915+105 for 3 epochs; 1994, 2023, and 2024, separated by about 30 years.
 
2) The displacement observed in the radio source tracing the X-ray binary is consistent with the accurate proper motions measured with the VLBA by Reid \& Miller-Jones (2023). 

3) The position angle and the inclination of the ejecta have significant changes in 2023. Between 1994 and 2023 the position angle 
increased in the counterclockwise sense by $\Delta PA = 24\rlap.^\circ 6 \pm 0\rlap.^\circ 9$. 
Assuming that the lobes were ejected at the same time and that $\beta$ is not too small
(i.e. $\beta \leq$ 0.5),
we conclude that between 1994 and 2023 the angle with the line of sight of the jet axis $i$ has increased by 
$17^\circ \pm 4^\circ$, to $i$ = $87^\circ \pm 3^\circ$, nearly perpendicular to the line of sight.

4) For inclination angles of the jets with respect to the plane of the sky of
$3^\circ \pm 3^\circ$, relativistic boosting and de-boosting should not be significant, and the displacements and
brightnesses of the Southern and Northern lobes in the observations of 2023 are about the same (Figure 1). 

5) Therefore, the VLA observations of 2023 confirm that the jets in GRS 1915+105 are antisymmetric ejections of relativistic twin plasma clouds, as assumed by Mirabel and Rodriguez (1994).
 
6) If the accretion disk is nearly perpendicular to the direction of the jets,
$i$ = $87^\circ \pm 3^\circ$,
 this geometry implies that the line of sight to the X-ray source in the observations of 2023 (Figure 1) is across the massive accretion disk. 

7) The historic periods of the radio flux oscillations of 20 to 40-min changed to a period of 8-min in 2023. From 
this change in period it is suggested that the semi-major axis of the accretion disk in 2023 could have
been reduced, implying an enhancement of the surface gas density of the disk and probably an increase 
in the accretion rate. 

8) Under this particular orientation of the line of sight, it was estimated an X-ray Compton-thick, dense-gas column density of 
$N_H > 3 \times 10^{23}~ cm^{-2}$ by Gandhi et al. (2025),
which explains the deep X-ray obscured state (Miller et al. 2020; Balakrishnan et al. 2021), and the 
exceedingly large mid-infrared luminosity observed with JWST in that epoch (Gandhi et al. 2025).

9) The amplitude of the abrupt change in the direction of the outflow axis of GRS 1915+105 in 2023 differs from the smaller changes 
previously observed by a factor of 3.4-$\sigma$, which is a significant change.

10) Figure 1c shows that after the unusual change of orientation of the radio jets in 2023, in 2024 the jets returned to the range of their 
historical orientations. This is an indication that the unusual orientation of the jets in 2023 was a transient event, rather than a change in the BH-spin. This return to
previous values is in agreement with the Bardeen-Petterson (1975) effect where the axis of a tilted disk around a Kerr black hole will finally transition to be aligned with the spin of the black hole.
 
\begin{acknowledgments}


L.F.R. acknowledges the financial support of PAPIIT - UNAM 
IN108324 and CONAHCyT 238631.
I.F.M. acknowledges the financial support of DAP-IRFU-CEA-France and thanks Andrew King for his thoughts on 
the possible causes of the changes reported here. We acknowledge the insightful comments of the anonymous
reviewer that improved the clarity of the paper.

\end{acknowledgments}

\end{document}